**Frequency Modulation of Spin-Transfer Oscillators**

M. R. Pufall, W. H. Rippard, S. Kaka, T. J. Silva, and S. E. Russek

*National Institute of Standards and Technology, Boulder, CO 80305*

**Abstract.**

Spin-polarized dc electric current flowing into a magnetic layer can induce precession of the magnetization at a frequency that depends on current. We show that addition of an ac current to this dc bias current results in a frequency modulated (FM) spectral output, generating sidebands spaced at the modulation frequency. The sideband amplitudes and shift of the center frequency with drive amplitude are in good agreement with a nonlinear FM model that takes into account the nonlinear frequency-current relation generally induced by spin transfer. Single-domain simulations show that ac current modulates the cone angle of the magnetization precession, in turn modulating the frequency via the demagnetizing field. These results are promising for communications and signal processing applications of spin-transfer oscillators.

## Introduction.

The excitation of magnetization precession by a high density dc electric current flowing perpendicularly through a multilayer spin-valve structure offers the possibility of both a new means of probing fundamental magnetic excitations and dynamics, and new microwave applications for magnetic thin-film devices.[1,2] In these systems, spin-polarized electric current flowing into a magnetic layer exerts a torque on the local magnetic moment if the spin polarization of the electric current and the magnetic moment are not parallel, an effect typically called spin-transfer.[3,4] In the point contact (nanocontact) structures discussed here, in which a 40 nm lithographic contact is made to a continuous spin-valve multilayer consisting of thick and thin magnetic layers separated by a nonmagnetic spacer layer, this torque excites magnetization precessional motion with a frequency ranging from 5 GHz to 40 GHz that is tunable via both current and applied field.[2] In addition, the excitation has a narrow (1.5 MHz - 50 MHz) room temperature linewidth, leading to quality factors $Q = f/\Delta f$ on the order of $10^4$ for these oscillators. The current-tunability and high $Q$ of these resonant structures suggest the possibility of their use in microwave signal processing applications.

Although spin-transfer-induced magnetization precession in nanocontacts was first predicted[4] and then subsequently observed,[2,5] many aspects of these observations were unanticipated, in particular the narrow spectral linewidth of the mode and the complicated dependence of the precession frequency on dc current. For example, point contact structures show both red and blue shifts and discrete jumps with increasing current, the details of which depend on the strength and direction of the applied field.[5] As a means of probing the stability of these modes, and as a basic test of their suitability for communicating information—since a pure tone carries no information—we have made measurements in which we add a small ac current perturbation to the dc bias current, and measure the resultant spectral output of the device as a function of the ac amplitude. We find that although the microwave signal generated by these oscillators arises from the precession of a magnetic moment in a highly nonlinear regime (i.e., large-angle precession), the devices nonetheless behave like conventional tunable oscillators. Nanocontact devices exhibit frequency modulation (FM) characteristics as a function of ac current amplitude that one would expect from a device with a nonlinear current-frequency transfer function, such as asymmetric sideband suppression, and shifting of the resonance center frequency. The modulation frequency $f_{mod}$ can vary widely for these devices, having been successfully modulated at frequencies up to $0.9f_{precession}$. We use a single-domain model that includes the Slonczewski spin-torque term[2,4] to show how the ac current affects the magnetization trajectory of the resonance, altering the cone angle of the precession and consequently changing the demagnetizing field, and thus the total effective field and frequency of precession.[6]

**Experiment.**

All measurements were made on structures consisting of point contacts made to continuous thin-film multilayers of the form Au(2.5nm)/Cu(1.5nm)/$Ni_{80}Fe_{20}$(5 nm)/Cu(5 nm)/Co90Fe10(20 nm)/Cu(50 nm)/$SiO_2$/Si, similar to those measured previously.[2,5] To excite the spin-transfer resonance, a dc current was sent through the device via the dc leg of a bias tee. A microwave power splitter was connected to the ac leg of the bias tee, allowing both injection of an ac signal into the device, and coupling of any generated high frequency signals out of the device, to be subsequently measured by a 50 GHz spectrum analyzer. As has been described in more detail previously,[2,5,7] the injected dc current induces precessional motion of the NiFe (free) layer of the spin valve above a critical current $I_c$. A change in the relative alignment of the two layers causes a change in the device resistance due to the giant magnetoresistance (GMR) effect, in turn creating an ac change in voltage that is then coupled out through the bias tee and power splitter.

**Results and Discussion.**

The spectral output of a typical resonance is shown in the inset to Figure 1. The frequency and amplitude of the resonance are functions of the applied field magnitude and direction, and the applied current.[5] The measurements reported here are for an applied field of 0.7 T at $80^{o}$ to the plane (shown schematically in Figure 1), a particular configuration chosen for the large microwave output amplitude, narrow linewidth, and smooth variation of the resonance frequency with current, as shown in Fig 1. This type of spin-transfer-induced precession has generally been observed over a range of applied field angles, with the details of frequency $f$ vs. current $I$ changing with field strength and direction.[5] The frequency increases roughly linearly over the range of currents from 6 mA to 8 mA, then increases at a more gradual rate for currents up to 9.5 mA.

To examine the effects of an ac current on the resonance, the device was then biased to a fixed current $I_0$ and an additional 40 MHz ac current was applied through the power splitter, generating a time-varying resonance frequency, i.e., frequency modulation. In Figure 2a, the spectral outputs at $I_0$ = 8.5 mA are shown for several input ac current amplitudes. The spectra generally show that with increasing modulation current amplitude $\Delta I$, more power is driven into sidebands positioned at $f = f_{center} \pm n \cdot 40$ MHz (the sideband order $n$ = 1, 2,…), with the specific sideband magnitudes depending on the variation of $f$ vs. $I$ in the vicinity of $I_0$. For example, at a bias point in the linear region of the $f$ vs. $I$ curve, such as $I_0$ = 7.5 mA, the upper and lower sidebands of a given order have approximately the same amplitude, whereas when biased at $I_0$

= 8.5 mA, a point in the curved region, the upper and lower sidebands have significantly different magnitudes, and vary differently with $\Delta I$ (Fig. 2a).

Multiple Lorentzians were simultaneously fit to the spectra for $I_0$ = 8.5 mA to determine the spectral peak positions and amplitudes, shown as functions of $\Delta I$ in Figs. 2b-d. The upper and lower sideband amplitudes of a given order vary in markedly different ways as a function of $\Delta I$. For example, for the first order sidebands ($f = f_0 \pm 40$ MHz, Fig. 2c), the upper sideband is larger in magnitude at a given $\Delta I$, and peaks at a $\Delta I$ higher than that for the lower sideband. In contrast, for the second order sidebands ($f = f_0 \pm 80$ MHz, Fig. 2d), the lower sideband is larger in magnitude for low $\Delta I$, with the upper sideband becoming the larger for $\Delta I > 0.75$ mA. Finally, the central peak (the "carrier" frequency $f_0$, see Fig. 2b) red shifts (decreases) significantly with $\Delta I$. As will be shown, these effects are due to the nonlinear shape of the $f$ vs. $I$ transfer curve in the neighborhood of 8.5 mA.

The general form for the output signal is $V(I,t) = \mathrm{Re}\{V_0 \exp(i\theta(I,t))\}$, where the phase angle is defined as $\theta(I,t) = 2\pi \int_0^t f(I(t'))dt'$. For a system with a linear $f$ vs. $I$ characteristic of slope $A$, if one assumes a current of the form $I(t) = I_0 + \Delta I \cos(2\pi f_{\mathrm{mod}} t)$, the phase angle becomes $\theta(I, t) = \omega_0 t + \beta \sin(\omega_{\mathrm{mod}} t)$, in which $\omega_0 = 2\pi f(I_0)$, $\omega_{\mathrm{mod}} = 2\pi f_{\mathrm{mod}}$, and the modulation factor $\beta \equiv f_{\mathrm{dev}}/f_{\mathrm{mod}} \equiv A\Delta I/f_{\mathrm{mod}}$. The resulting expression for $V(I,t)$ can be expanded in a Bessel series given by

$$V(I,t) = V_0 e^{i(\omega_0 t + \beta \sin(\omega_{\mathrm{mod}} t))} = V_0 e^{i\omega_0 t} \sum_{l=-\infty}^{\infty} J_l(\beta) e^{il\omega_{\mathrm{mod}} t}, \quad (1)$$

which shows that the amplitude of the $l$th sideband is proportional to $J_l(\beta)$ for linear FM.[8]

In the present case, $f$ vs. $I$ (see Fig. 1b) is a higher-order polynomial in $I$, and can be expanded as a Taylor series of order $n$ in the neighborhood of $I_0 = 8.5$ mA. The key point is that a sinsusoidal input current now results in a power series in $\Delta I\cos(\omega_{\mathrm{mod}}t)$ for $f(I,t)$. Since each factor $\cos^k(\omega_{\mathrm{mod}}t)$ can be expanded as a series in $\cos(m\omega_{\mathrm{mod}}t)$ ($m \leq k$), one finds that a series in $\sin(m\omega_{\mathrm{mod}}t)$ ($m \leq n$) now replaces the single sinusoidal term in Eq. 1. In addition, each even power in the cosine power series also contributes a term linear in $t$, terms which are the sources of the shift of the carrier (center) frequency.

With each harmonic of $\omega_{\mathrm{mod}}$ expanded as in Eq. 1, the expression for $V(I_0,\Delta I,t)$ is then a product of Bessel series, with the number of factors set by the order of the polynomial. For example, the third order expression is

$$V(I_0, \Delta I, t) = \mathrm{Re}\left\{ V_0 e^{iA_0 t} \sum_{l=-\infty}^{\infty} \sum_{m=-\infty}^{\infty} \sum_{p=-\infty}^{\infty} J_l(A_1) J_m(A_2) J_p(A_3) e^{i(l+2m+3p)\omega_{\mathrm{mod}} t} \right\} \quad (2)$$

in which the $A_{\mathrm{n}}$ are linear combinations of the Taylor coefficients, and $A_0$ is a sum of $\omega_0 = 2\pi f(I_0)$ and contributions from even powers in the Taylor expansion. The $A_{\mathrm{i}}$ are in general functions of $I_0$, $\Delta I$, and $\omega_{\mathrm{mod}}$, and are analogous to the modulation factor $\beta = f_{\mathrm{dev}}/f_{\mathrm{mod}}$ in linear FM. This sum has terms proportional to $\sin(A_0 t)$, $\sin(A_0 t \pm \omega_{\mathrm{mod}}\, t)$, $\sin(A_0 t \pm 2\omega_{\mathrm{mod}}\, t)$, $\sin(A_0 t \pm 3\omega_{\mathrm{mod}}\, t)$,… describing a carrier at a (shifted) frequency $A_0/2\pi$, plus sidebands at integer harmonics of $f_{\mathrm{mod}}$. A given sideband's amplitude is a sum of terms such that the sum $l + 2m + 3p$ is equal to the order of the sideband, i.e., ±1, ±2, etc. The indices $l$, $m$, $p$ can be either positive or negative, increasing the number of contributing terms to a given order, and producing the asymmetry in amplitudes between the upper and lower sidebands.

The computed center frequency shift and sideband amplitudes as functions of drive amplitude are shown as the solid lines in Figs. 2b-d, determined using a fifth order Taylor series for $f$ vs. $I$ around $I_0$ = 8.5 mA. As seen in Fig. 2, the above expressions accurately predict both the red shift of the center frequency, and reasonably predict the relative variations of the sideband amplitudes with $\Delta I$, up to a constant amplitude factor. The model describes the amplitude difference between the upper and lower sidebands of a given order, and also the crossover in their relative magnitudes (recall that for linear FM, the magnitudes of the upper and lower sidebands of a given order are equal.) The overall magnitudes of the calculated sidebands (but not the carrier frequency amplitude) are too large by a factor of 1.5, a factor that varied with bias point. This possibly results from nonlinearities in the *I-V* curve, or from nonlinear amplitude modulation effects not included in the theory. The amplitude of the output signal is not constant with $I$ (see coarse variation of amplitude in Fig. 1 inset), on average decreasing away from $I_0$, decreasing the amplitudes of the sidebands.[9]

It is worth emphasizing that these observed FM effects are not simply electrical, e.g., the result of signal mixing due to a nonlinear *I-V* relation, but rather correspond to periodic variations in the trajectory of the magnetization of the free layer of the nanocontact structure, detected via the GMR effect. As a heuristic method for understanding of the effects of low-frequency FM on the trajectory of the magnetization, simulations of current-induced dynamics in single-domain structures were performed. The simulations were based on integration of a Landau-Lifshitz-Gilbert equation, modified as described by Slonczewski[2,4] to include the effects of spin torque. As noted previously, these simulations give qualititative information on frequencies and trajectories, but fail to predict quantitative details such as the magnitudes of the critical current or the slope of the frequency vs. current.

Simulated trajectories of a 100 nm x 100 nm device for both zero and nonzero modulation amplitudes over one-half period of the modulation are shown in Fig. 3 for an applied field $\mu_0 H_{app}$ = 0.7 T at 80° to the film plane.[2] In this configuration, the simulations predict a roughly linear dependence of frequency on applied current. The trajectory for nonzero modulation has a larger width in the z-direction (perpendicular to the film plane). The modulation drives the magnetization periodically more into and out of the plane, expanding and contracting the cone of precession. This decreases and increases the demagnetizing field, which in turn modulates the net effective field and the precession frequency. The projection onto the y-z plane shows that the average value of the magnetization perpendicular to the plane oscillates at the modulation frequency.

These results support the notion that the large-angle precessional modes of the magnetization induced by spin transfer are stable and tolerant of significant (current-induced) perturbations. Current-driven modulation of the device produces frequency modulation effects (sidebands and frequency shifts) that are well-described by a standard FM model using the nonlinear current-frequency transfer curve determined experimentally. Modulation amplitudes up to a significant fraction of the dc bias current were used. These results, along with single-domain simulations, suggest that current-driven frequency modulation produces controlled variations of the magnetization trajectory, resulting in modulations of the effective field without destabilizing the trajectory.

**Figure Captions:**

**Figure 1:** Frequency of spin-transfer-induced precession in a nanocontact as a function of dc bias current. Schematic shows sample measurement geometry. Inset: Spectral output of nanocontact for several dc currents, showing variation of frequency and amplitude with current.

**Figures 2a-d:** Effects of injected ac current on spin-transfer resonance, at a dc current bias of $I_0$ = 8.5 mA, $f_{mod}$ = 40 MHz. Solid lines are sideband amplitudes, calculated as described in the text. 2a: Spectral output of device for several input modulation amplitudes $\Delta I$, showing sideband changes and shift of center frequency vs. $\Delta I$. 2b: Shift of center frequency vs. $\Delta I$. 2c: Variation of upper and lower first-order sideband amplitudes vs. $\Delta I$. 2d: Variation of upper and lower second-order sideband amplitudes vs. $\Delta I$. Right scales show calculated sideband amplitudes, left scale measured sideband amplitudes.

**Figure 3:** Plot of computed single-domain magnetization trajectory with (grayscale symbols) and without (light gray line) injected ac current modulation, over one modulation period. Color scale on trajectory denotes phase of ac modulation, from gray (ac current at minimum) to black (ac current at maximum). Axes are in units of the saturation magnetization. Trajectories also projected onto y-z plane to show spreading of the orbit in the z-direction with drive. Simulations shown are at $T$ = 0 K, to more easily see the trajectory. AC current values differ from measured values due to different slope of calculated $f$. vs. $I$ curve, and uncertainties in the absolute scaling of $I$ in the model. Inset: Fourier transforms of x-component of trajectories, showing frequency sidebands generated during modulation.

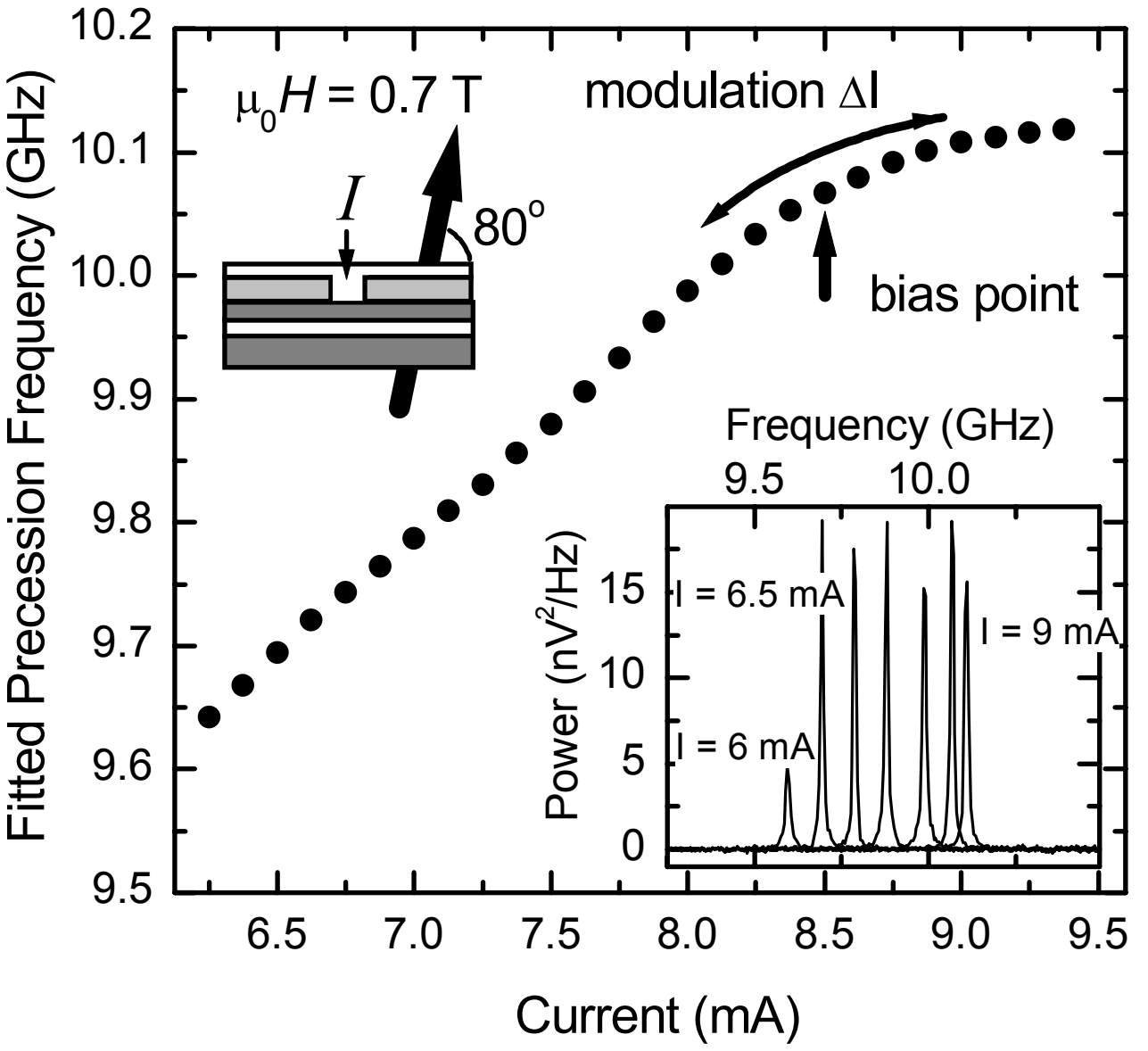

Figure 1, Pufall *et al.*

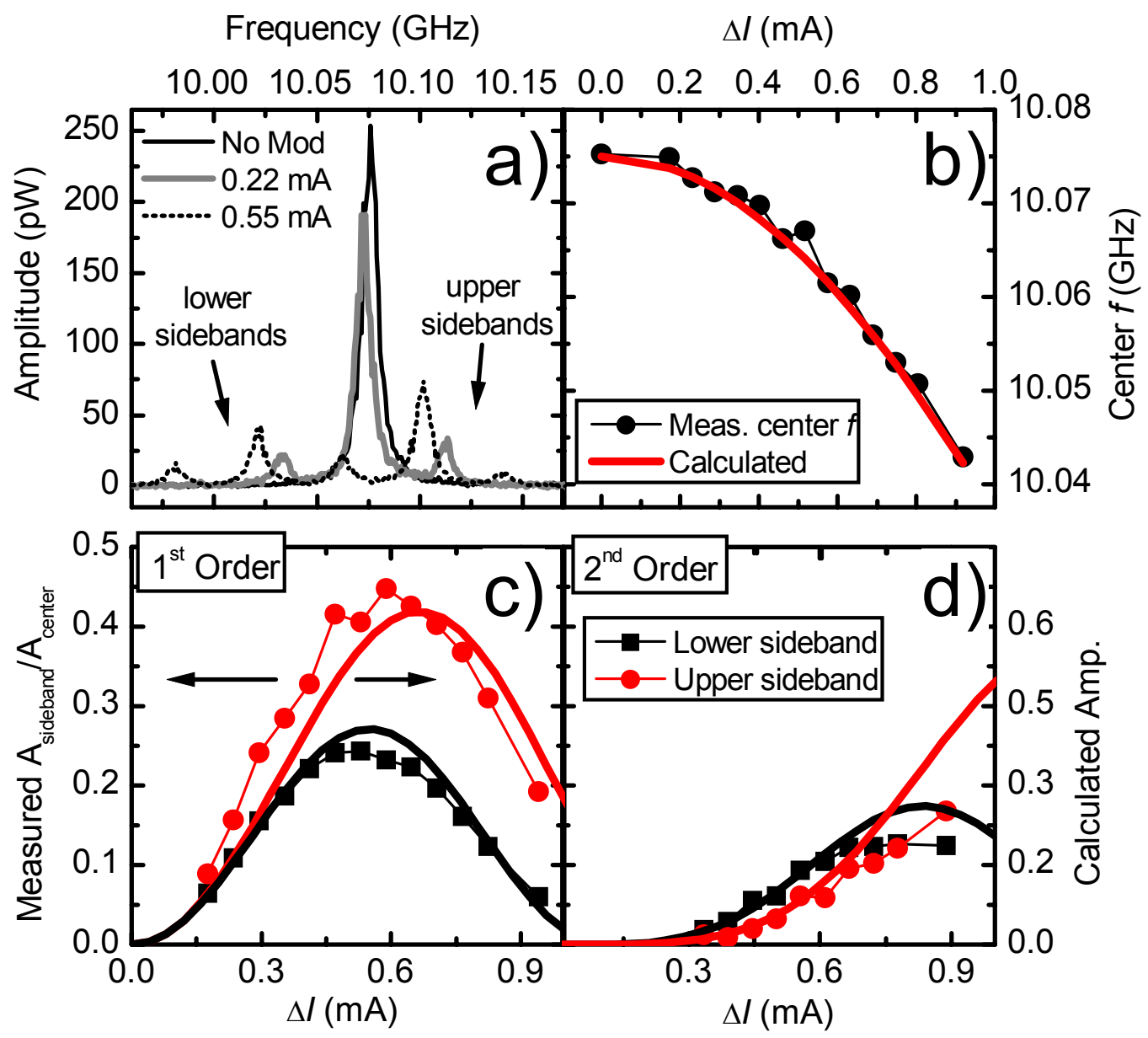

Figure 2, Pufall *et al.*

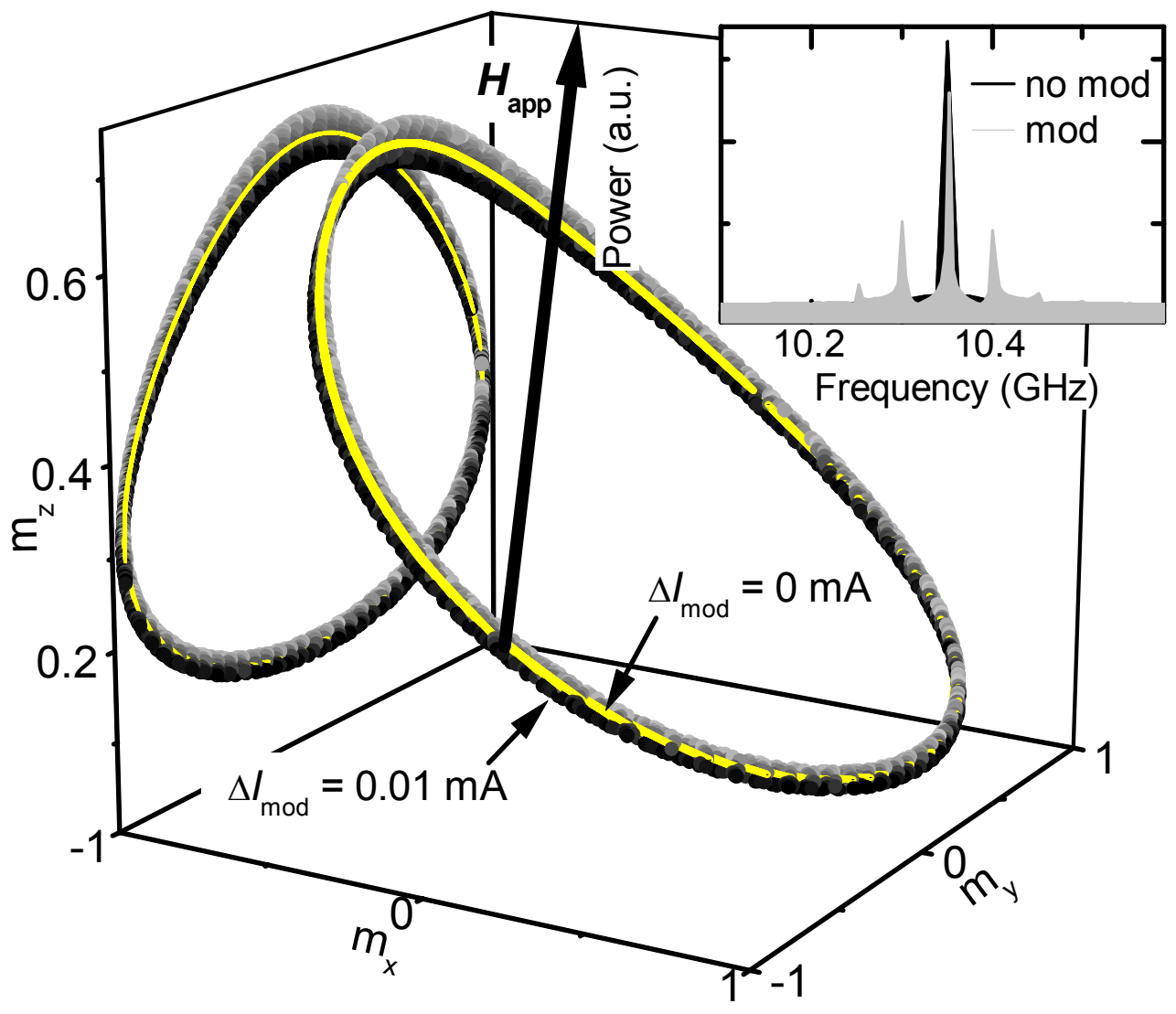

Figure 3., Pufall *et al.*